\documentclass{aa}
\usepackage[varg]{txfonts}

\begin{document}

\title{Power requirements for cosmic ray propagation models involving diffusive reacceleration; estimates and implications for the damping of interstellar turbulence.}
\titlerunning{Cosmic ray reacceleration power}

\author{Luke O'C. Drury\inst{1} \and Andrew W. Strong\inst{2}}

\institute{Dublin Institute for Advanced Studies, School of Cosmic Physics, 31 Fitzwilliam Place, Dublin 2, Ireland
\and
MPI f\"ur Extraterrestrische Physik, Postfach 1312, 85741 Garching, Germany}

\newcommand\Galprop{{\sc Galprop}}

\date{Received soon / accepted shortly thereafter}

\abstract{We make quantitative estimates of the power supplied to the Galactic cosmic ray population by second-order Fermi acceleration in the interstellar medium, or as it is usually termed in cosmic ray propagation studies, diffusive reacceleration.  Using recent results on the local interstellar spectrum from the Voyager missions we show that for parameter values, in particular the Alfv\'en speed,  typically used in propagation codes such as \Galprop\ to fit the B/C ratio, the power contributed by diffusive reacceleration is significant and can be of order 50\% of the total Galactic cosmic ray power.  The implications for the damping of interstellar turbulence are briefly considered.}


\maketitle

\section{Introduction}

In most discussions of cosmic ray propagation it is assumed that the cosmic rays are produced in discrete sources, generally taken to be supernova remnants, although there is increasing interest in and evidence for possible additional classes of sources, see e.g. the presentations in the recent conference organised by \citet{2014NuPhS.256....1T}, and 
the recent exciting evidence for acceleration of protons to $\rm PeV$ energies in the Galactic centre region \citep{2016Natur.531..476H}.
They then diffuse through the interstellar medium and an extended cosmic ray halo before escaping from the Galaxy.  This diffusion is not just in space, but also in momentum, if, as is inevitably the case, the magnetic fields that scatter the particles are not purely static but are in random motion with characteristic velocities of order the Alfv\'en speed.  This leads to a certain amount of second-order Fermi acceleration during propagation, an effect which for historical reasons is usually referred to as ``reacceleration''.  There has been a long and inconclusive debate as to the importance of reacceleration, but there is no doubt that the idea is attractive for a variety of reasons, and physically the effect must occur at some level. 
For clarity we note that the term ``reacceleration'' is also sometimes used to refer to the acceleration by SNR shocks of pre-existing ambient cosmic rays; in this paper we do not consider this shock reacceleration process and are exclusively concerned with diffusive reacceleration.

One of the strongest arguments in favour of diffusive reacceleration is that it allows a rather natural fit to the low energy Boron to Carbon data.  This was persuasively argued by \citet{1995ApJ...441..209H} who showed that with reacceleration the B/C data could be well fit with a single power-law dependence of the escape path-length as a function of rigidity, whereas in a leaky-box model one had to suppose an unnatural decrease in the path-length at low energies.  The fact that the inferred power-law dependence of the path-length also agreed with that predicted for Kolmogorov turbulence was seen as another positive feature of the model (although whether interstellar magnetic turbulence should be described by a Kolmogorov spectrum is another question). 

This reacceleration by interstellar turbulence is a potentially significant contribution to the total Galactic cosmic ray source power (and of course also a potentially significant damping term for interstellar turbulence).   This was qualitatively discussed in \cite{2014MNRAS.442.3010T} and presented at the ICRC by \cite{2015arXiv150802675O} on the basis of analytic estimates.  In this paper we seek to make quantitative estimates of the diffusive reacceleration power using both recent parametrisations of the local interstellar spectrum and the well-established cosmic ray propagation code \Galprop\ as first described in \cite{1998ApJ...509..212S}.  \Galprop\ has become the de facto reference model for cosmic ray propagation studies in our Galaxy and is thus an appropriate choice, but any other propagation model incorporating  diffusive reacceleration should give very similar results, e.g. the {\sc Dragon} code of \cite{2016arXiv160707886E}.  

The power transferred to the cosmic rays comes of course at the expense of the general turbulence in the interstellar medium so we also evaluate a damping process for interstellar turbulence.  An important aspect of this problem is that it depends (as we will see) sensitively on the low-energy form of the cosmic ray spectrum in interstellar space.  Until recently this was largely unknown, but we now have direct {\it in situ} observations from the Voyager spacecraft of the cosmic ray spectrum beyond the heliopause which we can use to constrain the local interstellar spectra at sub-relativistic energies \citep{2013Sci...341..150S}.  There are also useful constraints from the diffuse gamma-ray emission of the Galaxy \citep{2015arXiv150705006S} which are important as being independent of solar modulation and confirming the spectral break at GeV energies.

\section{Reacceleration power}

This section briefly reviews and expands on the main results of \cite{2014MNRAS.442.3010T}.  The basic equation underpinning all diffusion models of cosmic ray transport, going back at least to the classic monograph of  \cite{1964ocr..book.....G} is
\begin{equation}
{\partial f\over\partial t} = Q + 
\nabla\left(D_{xx}\nabla f\right) +  {1\over 4\pi p^2} {\partial\over\partial p}\left(4\pi p^2 D_{pp} {\partial f\over\partial p}\right) +....
\end{equation}
Here $f(x,p,t)$ is the isotropic part of the phase space density of a given species as a function of position $x$, scalar momentum $p$ and time $t$ and $Q(x, p, t)$ is a source term representing the initial production and acceleration of cosmic rays.  The spatial and momentum space diffusion coefficients are $D_{xx}$ and $D_{pp}$ respectively and will in general be functions of position, momentum, and time.  On heuristic grounds \citep{2014MNRAS.442.3010T} the two diffusion coefficients are related by
\begin{equation}
D_{xx}D_{pp} \approx {1\over 9} p^2 V_A^2.
\end{equation}

Detailed calculations \citep{1975MNRAS.172..557S,1975MNRAS.173..245S,1975MNRAS.173..255S},
see also equations 9.38 and 9.39 in \citet{1990acr..book.....B}, show that
\begin{equation}
D_{xx}D_{pp} =
p^2 V_A^2 
\left<1-\mu^2 \over \nu_+ +\nu_-\right>
\left<(1-\mu^2)\nu_+\nu_-\over\nu_+ + \nu_-\right>
\end{equation}
where the angle brackets denote an average over an isotropic pitch angle distribution $\mu$ being the pitch angle cosine,
\begin{equation}
\left<\phantom{1\over 2}\right> = \int_{-1}^{+1} {d\mu\over 2}
\end{equation}
and $\nu_\pm$ is the rate of pitch-angle scattering off forward and backward propagating Alfv\'en waves.  
If the scattering rates are taken to be constant and equal we recover the naive heuristic result, but in general the scattering rates should be functions of the pitch angle and related to the wave spectrum through the gyro-resonant condition.  This introduces a weak dependence on the shape of the wave spectrum, but not on its amplitude. Specifically, if we assume a power-law wave spectrum with $W(k)\propto k^{-a}$ where $\int W(k)\,dk$ is the total energy content of the waves and $k$ is the wave-number ($a=5/3$ being a Kolmogorov spectrum for example).  
\begin{equation}
\nu_\pm \approx {v\over r_{\rm g}} {k_{\rm res} W_\pm (k_{\rm res})\over B_0^2/2\mu_0}
\end{equation}
where $r_{\rm g} = p/eB_0$ is the particle gyroradius, $B_0$ the background magnetic field,  and the resonant condition is
\begin{equation}
\mu r_{\rm g} k_{\rm res} \approx 1.
\end{equation}

Considered as a function of pitch angle the scattering rates thus scale as
\begin{equation}
\nu_\pm \propto \mu^{a-1}
\end{equation}
so that (assuming $\nu_+= \nu_-$)
\begin{eqnarray}
&&\left<1-\mu^2 \over \nu_+ +\nu_-\right>
\left<(1-\mu^2)\nu_+\nu_-\over\nu_+ + \nu_-\right>\nonumber\\
&=&
{1\over 4} \int_0^1 (1-\mu^2)\mu^{1-a}d\mu \int_0^1(1-\mu^2)\mu^{a-1}d\mu\nonumber\\
&=&
{1\over 4} \left({1\over 2-a} - {1\over 4-a}\right)
\left({1\over a} - {1\over 2+a}\right), \quad 0<a<2, \nonumber\\
&=& {1\over a(4-a)(4-a^2)}.
\end{eqnarray}
Thus more generally
\begin{equation}
D_{xx}D_{pp} =
p^2 V_A^2 {1\over a(4-a)(4-a^2)}
\end{equation}
which agrees with the simple heuristic estimate for $a=1$.  

However if we hold $\mu$ fixed and look at the scattering rate as a function of particle momentum or gyro-radius, then
\begin{equation}
\nu_\pm \propto v r_{\rm g}^{-2} W(k_{\rm res}) \propto v r_{\rm g}^{(a-2)}
\end{equation}
and thus
\begin{equation}
D_{xx} \propto v^2 \left<(1-\mu^2){1\over \nu_+ + \nu_-}\right> \propto v \left(p\over e\right)^{(2-a)}
\end{equation}
which is the commonly used form, $D\propto vR^{\delta}$ or velocity times a power-law in particle rigidity (momentum per charge, which is directly proportional to the gyro-radius for fixed magnetic field).  The power-law index of the rigidity dependence of the spatial diffusion coefficient, $\delta$ is thus related to that of the wave spectrum by
\begin{equation}
\delta = 2 - a, \qquad a = 2 -\delta
\end{equation}
and as is easily verified
\begin{equation}
{1\over a(4-a)(4-a^2)} = {1\over \delta(4-\delta)(4-\delta^2)}.
\end{equation}
In fact the denominator can be written
\begin{equation}
a(4-a)(4-a^2) = 9 - 10(a-1)^2 + (a-1)^4
\end{equation}
which shows that it is symmetric about and has a local maximum at $a=1$.
Thus within this simple quasilinear theory of particle scattering we have
\begin{equation}
D_{xx}D_{pp} = p^2 V_A^2 {1\over \delta(4-\delta)(4-\delta^2)}\ge {1\over 9} p^2 V_A^2 
\end{equation}
which is essentially the form used by \Galprop, see equation (1) in \citet{1998ApJ...509..212S} derived from \citet{1994ApJ...431..705S}, if their factor $4/3w$ is set to unity (the inclusion of $w$ appears formally incorrect, but as it is treated as a constant of order unity it makes no difference to the results).  It is interesting to note that for equal intensities of forward and backward propagating waves the naive scattering estimate corresponds to the minimum possible level of reacceleration, but of course lower values can be obtained with anisotropic wave fields.

For convenience we will from now on write
\begin{equation}
D_{pp} = \vartheta p^2 V_A^2 {1\over D_{xx}}.
\end{equation}
For a Kolmogorov spectrum with $\delta= 1/3$ we have $\vartheta = 81/385 \approx 0.21$ and the largest it can be is 
for $\delta = 1$ where we have $\vartheta = 1/9 \approx 0.11$; the results of \citet{1998ApJ...509..212S} can be obtained by setting\begin{equation}
\vartheta = {4\over 3 \delta(4-\delta^2)(4-\delta) w}.
\end{equation}

The local reacceleration power density (energy per unit time and per unit volume) is shown in  \cite{2014MNRAS.442.3010T}, assuming very reasonable regularity conditions on the particle distribution function, to be given by the integral over the particle spectrum,
\begin{equation}
P_R= \int_0^\infty 4\pi p^2 f {1\over p^2} {\partial\over\partial p} \left(p^2 D_{pp} v\right) \, dp
\end{equation}
which can be written, if $D_{pp}$ is expressed in terms of $D_{xx}$ using the above relation, as
\begin{equation}
P_R =  \int_0^\infty 
4\pi p^2 f \left (\vartheta V_A^2 p v\over D_{xx}\right) \left[ 4 + {\partial\ln(v/D_{xx})\over\partial\ln p}\right]\, dp.
\end{equation}
If we parametrise the spatial diffusion in the usual fashion as
\begin{equation}
D_{xx}=D_0 \left(v\over c\right) \left(p\over m c\right)^\delta
\end{equation}
where $m$ is the particle mass and $c$ the speed of light
this can be written in the useful and rather transparent form
\begin{equation}
P_R = \vartheta (4-\delta) {V_A^2\over D_0} m c^2 \int 4\pi p^2 f \left(p\over mc\right)^{1-\delta} dp
\label{P_Rfinal}
\end{equation}
which gives the local reacceleration power density in terms of a simple integral over the spectrum.  
The formula is interesting in that the term $V_A^2/D$ is a rate, $mc^2$ is just the particle rest mass energy, and the integral is essentially just the total number density of cosmic ray particles biased by a power-law factor with exponent $1-\delta$.  Thus the energy transfer is more or less as if all particles were acquiring their own rest mass energy on a time scale of $D/V_A^2$.  Integrating this over the entire Galaxy then gives the total power transferred to the cosmic ray population from Alfv\'enic turbulence.

\subsection{Numerical estimates using Voyager data}

Numerical evaluation of this integral requires knowledge of the low-energy part of the interstellar cosmic ray spectrum, as noted in \cite{2014MNRAS.442.3010T}, and until recently this was quite poorly constrained.  However the situation has dramatically changed with the recent passage of the Voyager spacecraft into the local interstellar medium which allows us for the first time to have access to {\it in situ} measurements of the low-energy cosmic ray flux just outside the heliosphere
\citep{2014BrJPh..44..581P}.  A useful analytic fit to the local interstellar flux of protons, $J_{\rm LIS}$, has been given by
\citet{2015ApJ...815..119V} in the form
\begin{equation}
j_{\rm LIS} = 2700.0 {T^{1.12}\over\beta^2}\left(T+0.67\over 1.67\right)^{-3.93}\,\rm m^{-2} s^{-1} sr^{-1} GeV^{-1}
\end{equation}
where $T$ is the kinetic energy in units of GeV and $\beta = v/c$ is the dimensionless particle speed.  
Noting that $dT = v\,dp$ and that $p^2 f(p) v \,dp = j(T)\, dT$ where $p$ is particle momentum and $f$ is the phase space density we can convert this flux to a local interstellar phase space density via
\begin{equation}
f(p) = p^{-2} j(T)
\end{equation}

\begin{figure}[htbp]
\begin{center}
\includegraphics[width=0.5\textwidth] {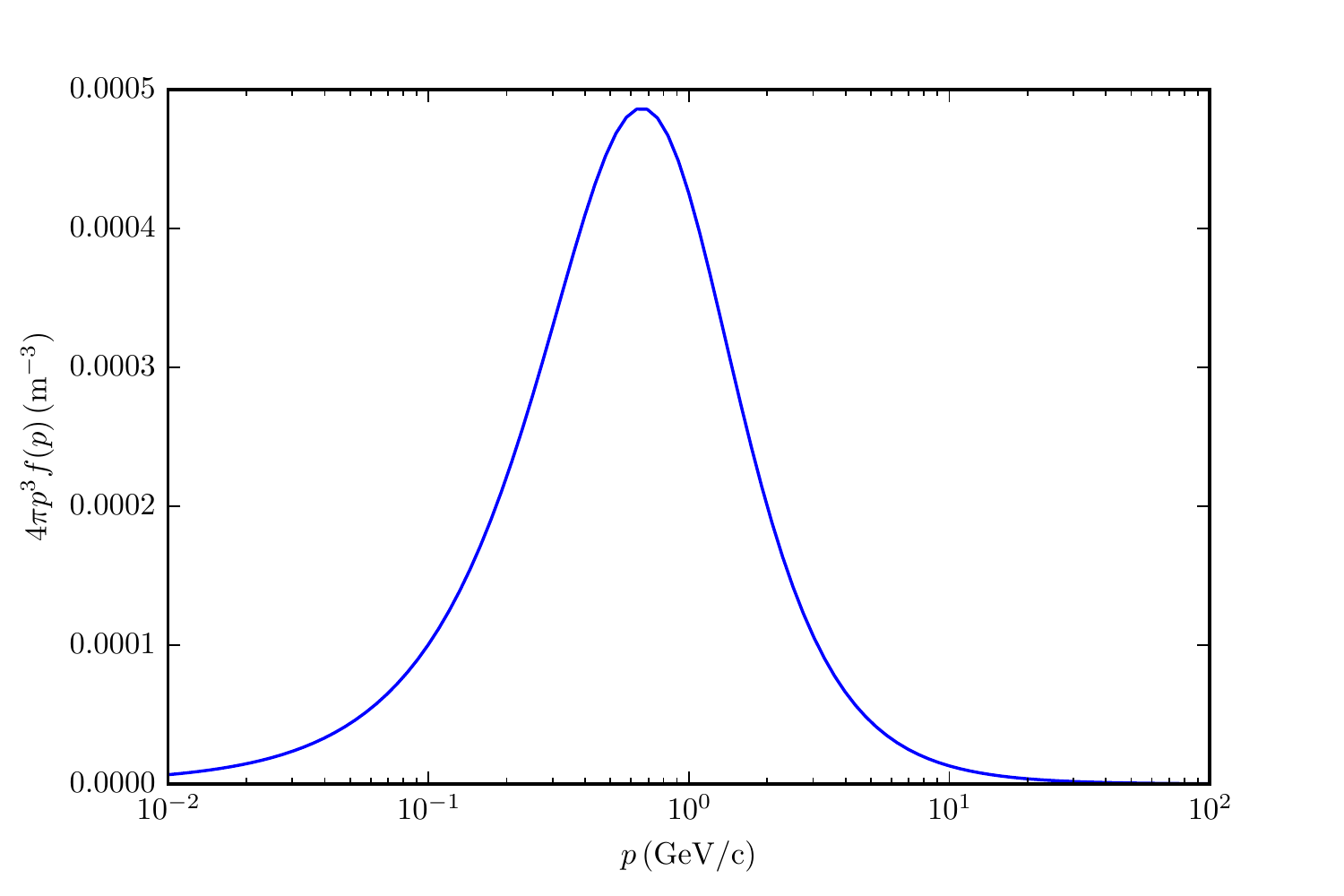}
\caption{The local interstellar proton number spectrum from \citet{2015ApJ...815..119V}}
\label{fig1}
\end{center}
\end{figure}

\begin{figure}[htbp]
\begin{center}
\includegraphics[width=0.5\textwidth] {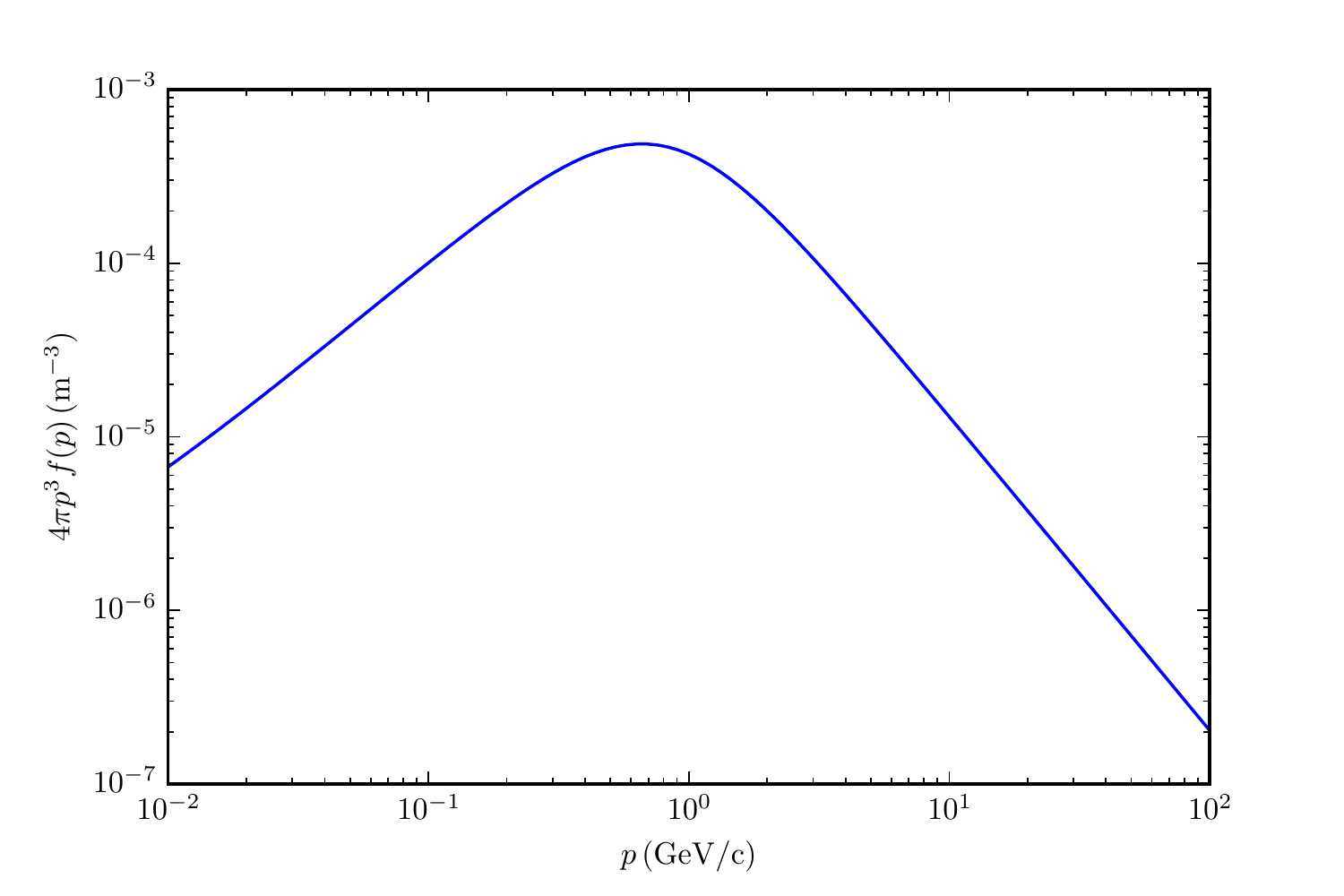}
\caption{The local interstellar proton number spectrum from \citet{2015ApJ...815..119V} in a more conventional log log representation showing the power-law tails}
\label{fig2}
\end{center}
\end{figure}

In Fig. \ref{fig1}
we show the corresponding spectrum plotted slightly unconventionally as a differential number spectrum in log momentum; we plot the number of protons per logarithmic interval in momentum, $4\pi p^3 f(p)$ noting that the total number density of particles is just the area under this curve,
\begin{equation}
n = \int 4\pi p^3 f(p) \,d \ln(p).
\end{equation}
Presenting the data in this way brings out just how sharply peaked the local interstellar cosmic ray proton spectrum is 
around $1\,\rm GeV$.  On a more conventional log log plot one can see that there are power-law tails extending to both lower and higher energies, see Fig. \ref{fig2}, but by far the dominant population, at least by number, are the mildly relativistic ones.
It is interesting to note that this spectrum corresponds to a total number density of $1.2\times 10^{-3} $ cosmic ray protons  per cubic meter locally, an energy density of $0.7 \,\rm eV \, cm^{-3}$
and a pressure of $0.33\,\rm eV\,cm^{-3}$.

\begin{figure}[htbp]
\begin{center}
\includegraphics[width=0.5\textwidth] {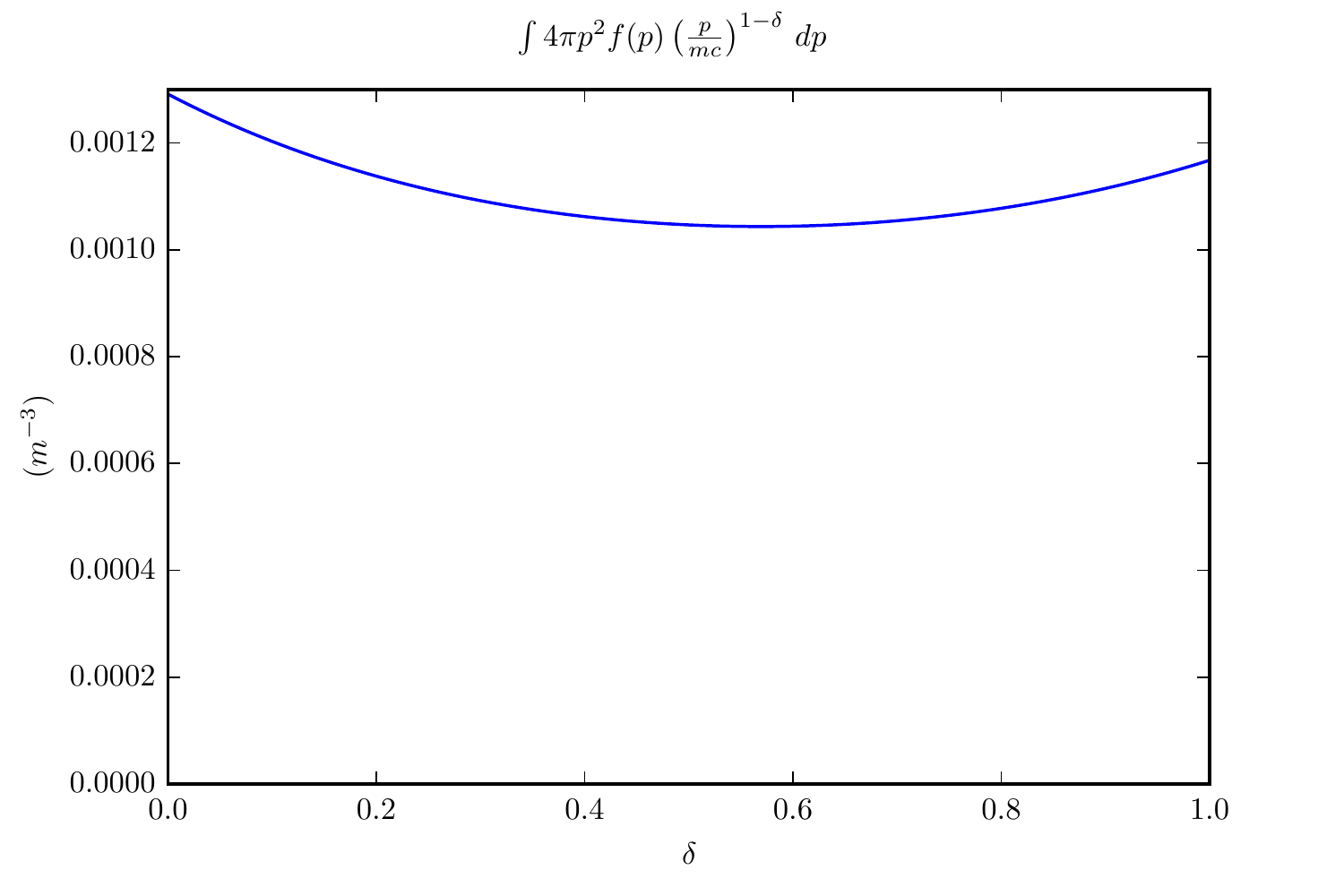}
\caption{The integral of $(p/mc)^{1-\delta}$ weighted by the local interstellar proton number spectrum from \citet{2015ApJ...815..119V} as a function of $\delta$ - note that the dimensions are $\rm m^{-3}$.  The value for $\delta =1$ is just the cosmic ray number density.}
\label{fig3}
\end{center}
\end{figure}

\begin{figure}[htbp]
\begin{center}
\includegraphics[width=0.5\textwidth] {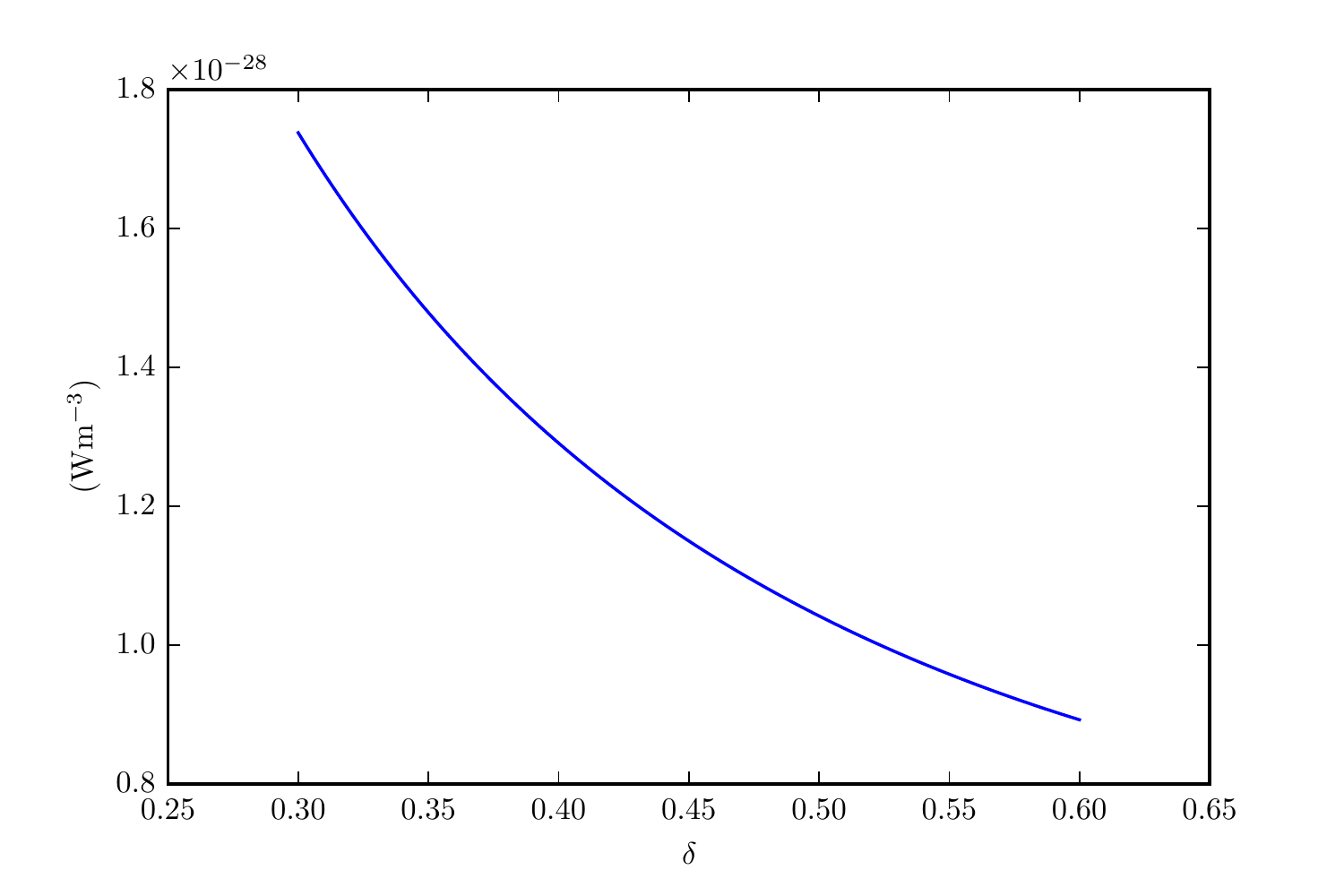}
\caption{The local reacceleration power density from Eq. \ref{P_Rfinal}, the Vos and Potgieter local interstellar spectrum, $V_A = 30 \, \rm km\,s^{-1}$ and $D_0 = 10^{28} \rm cm^2s^{-1}$}.
\label{fig4}
\end{center}
\end{figure}

We can now evaluate the integral in Eq. \ref{P_Rfinal}  over the local interstellar spectrum as parametrized by Vos and Potgieter.  The result is shown in Fig \ref{fig3} and shows a very weak dependence on $\delta$.   In Fig. \ref{fig4} we show the corresponding power density from Eq. \ref{P_Rfinal} using
\begin{equation}
\vartheta = {4\over 3 \delta(4-\delta^2)(4-\delta)},
\end{equation}
and canonical values
\begin{equation}
V_A = 30{\,\rm km\,s^{-1}}\qquad D_0 = 10^{28} \,\rm cm^2s^{-1}.
\end{equation}
We only plot the physically relevant range $0.3<\delta<0.6$ where we see that the power density is \begin{equation}
P_R \approx 1.3\pm0.4 \times 10^{-28} \,\rm W\, m^{-3} = 1.3\pm0.4 \times 10^{-27} \,\rm erg\, cm^{-3}
\end{equation}
if we now naively take the effective volume for the Galaxy to be $4\times 10^{61} \rm\, m^3$ then the total power input from diffusive reacceleration can be estimated as $5\times 10^{33} \,\rm W = 5\times 10^{40} \,\rm erg\, s^{-1}$.  
This should be compared with the canonical estimate of the total cosmic ray Galactic luminosity of order $10^{34}\,\rm W$ or $10^{41}\,\rm erg\, s^{-1}$ (see discussion in \citet{2014arXiv1412.1376O} where it is noted that there is probably half a decade of systematic uncertainty in this estimate either way).  Thus we have the, at first sight rather remarkable, result that diffusive reacceleration may be contributing as much as 50\% of the total cosmic ray luminosity of the Galaxy!

It is worth noting that if there is still some residual solar modulation beyond the heliopause, the local interstellar spectrum would have even more low energy particles and the diffusive reacceleration power would increase. Similarly, it is hard to believe that the local low-density bubble in which the solar system is located contains a significantly higher density of low-energy cosmic rays than the general Galactic average at the Sun's distance from the Galactic centre; if anything, adiabatic expansion of the hot bubble would imply a slightly lower value.  The ionization rates inferred from interstellar chemistry in molecular clouds also seem to indicate that the solar system cosmic ray environment is quite representative of the local Galaxy, as does the diffuse gamma-ray emission of the Galaxy (although at higher energies), see e.g. \citet{2015ARA&A..53..199G}.

\subsection{Estimates using  GALPROP}
As a cross-check on the above we now use the numerical CR-propagation package \Galprop\footnote{Current version available at https://sourceforge.net/projects/galprop},
using  parameters which reproduce B/C with reacceleration, and then with the same parameters but without the reacceleration term, so that the difference in energy content is due to the reacceleration process\footnote{\Galprop\  usually normalizes to the local proton spectrum, but in this case we want to study the effect of reacceleration so the normalization is not performed for the non-reacceleration runs.}
The total CR proton energy content of the Galaxy is computed by integration over momentum and volume.

We base the calculation on the model z04LMS, which has reacceleration and a CR halo height of 4 kpc and $V_A = 30\ \,\rm km \,  s^{-1}$ \cite{2010ApJ...722L..58S}. This model has typical parameter values which reproduce B/C in reacceleration models (see also \cite{2011ApJ...729..106T,2016ApJ...824...16J}).

The CR proton energy content of the Galaxy is $(8.1,\ 6.4)\times 10^{55}$erg with and without reacceleration respectively.  Hence $1.7 \times 10^{55}$ erg results from reacceleration, or $\approx$20\% of of the total; stated another way, the original energy injected by sources is boosted by $\approx$25\%.

The luminosity of the Galaxy can be estimated from the total energy content using a value for the residence time of CR in the Galaxy.
The  total proton luminosity for our reacceleration model\footnote{based on the total source luminosity input in \Galprop} is  $8\times 10^{40}$   erg   s$^{-1}$ \cite{2010ApJ...722L..58S}, so the effective CR residence time is $\approx3\times10^7$ yr. Hence the proton luminosity is  $8.0,\ 6.3 \times 10^{40}\,\rm erg\,  s^{-1}$  with and without reacceleration respectively.
This is consistent with the analytical estimates in Section 2.1.
Note that the relative values quoted above are independent of the assumed residence time.

\section{Conclusions and implications for interstellar turbulence}

Our estimates of the reacceleration power are necessarily approximate, but suffice to demonstrate that, for parameters chosen to reproduce B/C and commonly used in propagation calculations,
the energy input via reacceleration  is of order a quarter to half the total energy input.
SNRs are then not ``{\it the} sources of CR'' in such models, at least not in the conventional sense!
In retrospect this is perhaps not as surprising as it first seems.  If the bulk of the energy resides in mildly relativistic and sub-relativistic particles, and if the spectrum in this region is to be significantly modified by reacceleration, then the energy input must be significant.  

An interesting consequence is that the cosmic rays in such models must also damp interstellar turbulence and it is interesting to ask whether this is physically plausible and significant. Energetically it is not impossible.
The main energy input into interstellar turbulence is generally taken to be the mechanical energy from expanding SNR shells
and there is thus enough power input, but at quite large scales \citep{2004ARA&A..42..211E, 2004ARA&A..42..275S}. The turbulence is normally assumed to be dissipated by non-linear cascading to high wave-number modes and the question is whether the cosmic rays can extract enough energy from the cascade at the scales which scatter mildly relativistic particles before thermal dissipation takes over.  This is far from obvious, but if it is the case then the cosmic rays may define the inner scale of the interstellar turbulence.  SNRs would then remain the ultimate engine driving cosmic ray acceleration, but through two channels;  a direct one involving shock acceleration, and an indirect one mediated by interstellar turbulence.

Meanwhile it is essential to consider alternative explanations of the peak in the energy-dependence B/C, in particular convection gives a fairly natural mechanism \citep[e.g.][]{2016arXiv160706093K} and 
there is evidence for a Galactic wind \citep[e.g.][]{2012EAS....56...73E}.  The main message of this work is that behind the reacceleration term in the propagation codes there is a significant impact on the total energy budget of both the cosmic rays and the interstellar turbulence.  It is not a free parameter than one can tune at will without physical consequences.

\bibpunct{(}{)}{;}{a}{}{,} 
\bibliographystyle{aa}
\bibliography{QERP}

\begin{thebibliography}{26}
\expandafter\ifx\csname natexlab\endcsname\relax\def\natexlab#1{#1}\fi

\bibitem[{{Berezinskii} {et~al.}(1990){Berezinskii}, {Bulanov}, {Dogiel}, \&
  {Ptuskin}}]{1990acr..book.....B}
{Berezinskii}, V.~S., {Bulanov}, S.~V., {Dogiel}, V.~A., \& {Ptuskin}, V.~S.
  1990, {Astrophysics of cosmic rays} (North Holland, Amsterdam)

\bibitem[{{Drury}(2014)}]{2014arXiv1412.1376O}
{Drury}, L.~O. 2014, ArXiv e-prints [\eprint[arXiv]{1412.1376}]

\bibitem[{{Drury} \& {Strong}(2015)}]{2015arXiv150802675O}
{Drury}, L.~O. \& {Strong}, A.~W. 2015, ArXiv e-prints
  [\eprint[arXiv]{1508.02675}]

\bibitem[{{Elmegreen} \& {Scalo}(2004)}]{2004ARA&A..42..211E}
{Elmegreen}, B.~G. \& {Scalo}, J. 2004, \araa, 42, 211

\bibitem[{{Everett} {et~al.}(2012){Everett}, {Zweibel}, {Benjamin}, {McCammon},
  {Schiller}, {Rocks}, \& {Gallagher}}]{2012EAS....56...73E}
{Everett}, J., {Zweibel}, E., {Benjamin}, B., {et~al.} 2012, in EAS
  Publications Series, Vol.~56, EAS Publications Series, ed. M.~A. {de
  Avillez}, 73--76

\bibitem[{{Evoli} {et~al.}(2016){Evoli}, {Gaggero}, {Vittino}, {Di Bernardo},
  {Di Mauro}, {Ligorini}, {Ullio}, \& {Grasso}}]{2016arXiv160707886E}
{Evoli}, C., {Gaggero}, D., {Vittino}, A., {et~al.} 2016, ArXiv e-prints
  [\eprint[arXiv]{1607.07886}]

\bibitem[{{Ginzburg} \& {Syrovatskii}(1964)}]{1964ocr..book.....G}
{Ginzburg}, V.~L. \& {Syrovatskii}, S.~I. 1964, {The Origin of Cosmic Rays},
  authorised english translation edn. (Oxford: Pergamon Press)

\bibitem[{{Grenier} {et~al.}(2015){Grenier}, {Black}, \&
  {Strong}}]{2015ARA&A..53..199G}
{Grenier}, I.~A., {Black}, J.~H., \& {Strong}, A.~W. 2015, \araa, 53, 199

\bibitem[{{Heinbach} \& {Simon}(1995)}]{1995ApJ...441..209H}
{Heinbach}, U. \& {Simon}, M. 1995, \apj, 441, 209

\bibitem[{{HESS Collaboration} {et~al.}(2016){HESS Collaboration},
  {Abramowski}, {Aharonian}, {Benkhali}, {Akhperjanian}, {Ang{\"u}ner},
  {Backes}, {Balzer}, {Becherini}, {Tjus}, \& et~al.}]{2016Natur.531..476H}
{HESS Collaboration}, {Abramowski}, A., {Aharonian}, F., {et~al.} 2016, \nat,
  531, 476

\bibitem[{{J{\'o}hannesson} {et~al.}(2016){J{\'o}hannesson}, {Ruiz de Austri},
  {Vincent}, {Moskalenko}, {Orlando}, {Porter}, {Strong}, {Trotta}, {Feroz},
  {Graff}, \& {Hobson}}]{2016ApJ...824...16J}
{J{\'o}hannesson}, G., {Ruiz de Austri}, R., {Vincent}, A.~C., {et~al.} 2016,
  \apj, 824, 16

\bibitem[{{Korsmeier} \& {Cuoco}(2016)}]{2016arXiv160706093K}
{Korsmeier}, M. \& {Cuoco}, A. 2016, ArXiv e-prints
  [\eprint[arXiv]{1607.06093}]

\bibitem[{{Potgieter}(2014)}]{2014BrJPh..44..581P}
{Potgieter}, M. 2014, Brazilian Journal of Physics, 44, 581

\bibitem[{{Scalo} \& {Elmegreen}(2004)}]{2004ARA&A..42..275S}
{Scalo}, J. \& {Elmegreen}, B.~G. 2004, \araa, 42, 275

\bibitem[{{Seo} \& {Ptuskin}(1994)}]{1994ApJ...431..705S}
{Seo}, E.~S. \& {Ptuskin}, V.~S. 1994, \apj, 431, 705

\bibitem[{{Skilling}(1975{\natexlab{a}})}]{1975MNRAS.172..557S}
{Skilling}, J. 1975{\natexlab{a}}, \mnras, 172, 557

\bibitem[{{Skilling}(1975{\natexlab{b}})}]{1975MNRAS.173..245S}
{Skilling}, J. 1975{\natexlab{b}}, \mnras, 173, 245

\bibitem[{{Skilling}(1975{\natexlab{c}})}]{1975MNRAS.173..255S}
{Skilling}, J. 1975{\natexlab{c}}, \mnras, 173, 255

\bibitem[{{Stone} {et~al.}(2013){Stone}, {Cummings}, {McDonald}, {Heikkila},
  {Lal}, \& {Webber}}]{2013Sci...341..150S}
{Stone}, E.~C., {Cummings}, A.~C., {McDonald}, F.~B., {et~al.} 2013, Science,
  341, 150

\bibitem[{{Strong}(2015)}]{2015arXiv150705006S}
{Strong}, A.~W. 2015, ArXiv e-prints [\eprint[arXiv]{1507.05006}]

\bibitem[{{Strong} \& {Moskalenko}(1998)}]{1998ApJ...509..212S}
{Strong}, A.~W. \& {Moskalenko}, I.~V. 1998, \apj, 509, 212

\bibitem[{{Strong} {et~al.}(2010){Strong}, {Porter}, {Digel},
  {J{\'o}hannesson}, {Martin}, {Moskalenko}, {Murphy}, \&
  {Orlando}}]{2010ApJ...722L..58S}
{Strong}, A.~W., {Porter}, T.~A., {Digel}, S.~W., {et~al.} 2010, \apjl, 722,
  L58

\bibitem[{{Thornbury} \& {Drury}(2014)}]{2014MNRAS.442.3010T}
{Thornbury}, A. \& {Drury}, L.~O. 2014, \mnras, 442, 3010

\bibitem[{{Tibolla} \& {Drury}(2014)}]{2014NuPhS.256....1T}
{Tibolla}, O. \& {Drury}, L. 2014, Nuclear Physics B Proceedings Supplements,
  256, 1

\bibitem[{{Trotta} {et~al.}(2011){Trotta}, {J{\'o}hannesson}, {Moskalenko},
  {Porter}, {Ruiz de Austri}, \& {Strong}}]{2011ApJ...729..106T}
{Trotta}, R., {J{\'o}hannesson}, G., {Moskalenko}, I.~V., {et~al.} 2011, \apj,
  729, 106

\bibitem[{{Vos} \& {Potgieter}(2015)}]{2015ApJ...815..119V}
{Vos}, E.~E. \& {Potgieter}, M.~S. 2015, \apj, 815, 119

\end{thebibliography}

\end{document}